\documentclass[12pt]{article}
\usepackage{makeidx}
\usepackage{amsfonts}
\usepackage{amsmath}
\usepackage{epsfig}
\usepackage{latexsym}
\usepackage{amssymb}
\usepackage{graphicx}

\def\1{{\bf 1}}

\begin{document}

\title{Quantum signal+noise models: beyond i.i.d.\thanks{%
The paper is an extended version of author's lecture at \textquotedblleft
B.I.I.D.2016\textquotedblright\ workshop. }}
\author{A. S. Holevo \\
Steklov Mathematical Institute \\
of Russian Academy of Sciences, Moscow}
\date{}
\maketitle

\begin{abstract}
Recently, the Gaussian optimizer conjecture in quantum information theory
was confirmed for bosonic Gaussian gauge-covariant or contravariant
channels. These results use the \emph{i.i.d.} model of the quantum noise.

In this paper we consider quantum Gaussian signal+noise model with \emph{%
time-continuous stationary coloured noise}. A proof of the coding theorem
for the classical capacity of quantum broadband gauge-covariant Gaussian
channels is proposed. We also discuss and compare the ``broadband'' and the
``bandpass'' models of time-continuous time-continuous stationary coloured
noise.
\end{abstract}

\section{Introduction}

Recently, the Gaussian optimizer conjecture in quantum information theory
was confirmed for bosonic Gaussian gauge-covariant or contravariant channels
including phase-insensitive channels such as attenuators, amplifiers and
additive classical noise channels \cite{ghg}. It is shown that the classical
capacity of these channels under the input energy constraint is additive and
achieved by Gaussian encodings. These results use the \emph{i.i.d.} model of
quantum noise.

In this paper we consider a quantum Gaussian signal+noise model with \emph{%
time-continuous stationary coloured noise}. In this context we propose a
proof of coding theorem for the classical capacity of quantum broadband
gauge-covariant Gaussian channels. We also discuss and compare the
\textquotedblleft broadband\textquotedblright\ and the \textquotedblleft
bandpass\textquotedblright\ models of time-continuous stationary noise. It
is well-known that in the classical case a rigorous treatment of Gaussian
channels with time-continuous stationary coloured noise requires some
advanced mathematical tools such as the spectral theory for integral
operators with continuous symmetric kernels and Kac-Merdock-Sz\"{o}ge
theorem (see ch. 8 of \cite{xgal}). The present paper makes a step towards
achieving similar goal in the quantum case, where additional difficulties
due to the symplectic structure related to commutation properties of
observed processes arise.

There were several previous works where such problems were considered for
different special cases, with different degree of justification. In this
paper we rely upon the proof of coding theorem for a model \textquotedblleft
classical signal+quantum Gaussian noise\textquotedblright\ involving the
Planck spectrum, given in \cite{hol98}. In the paper of V. Giovannetti, S.
Lloyd, L. Maccone, P.W. Shor \cite{glm} the authors considered a broadband
pure-loss channel by formal passage from discrete to continuous spectrum and
demonstrated numerical solutions for the capacities $C,\,C_{ea},\,Q.$ The
paper of G. De Palma, A. Mari, V. Giovannetti \cite{pmg} was devoted to a
rigorous treatment of discrete time, Markov memory model, with flat noise
spectrum. Recently B. R. Bardhan, J. H. Shapiro \cite{bs} studied a
narrowband approximation for phase-insensitive time-invariant channels using
the result of \cite{ghg}.


The classical AGWN model is given by the equation%
\begin{equation}
Y_{k}=X_{k}+Z_{k};\quad k=1,\dots ,n  \label{awgn}
\end{equation}%
where $Z_{k}\backsim \mathcal{N}$ $(0,N)$ are real Gaussian i.i.d. random
variables representing the noise and the signal sequence $X_{k}$ is subject
to the energy constraint
\begin{equation*}
n^{-1}(X_{1}^{2}+\dots +X_{n}^{2})\leq E.
\end{equation*}%
\newline
The asymptotic ($n\rightarrow \infty $) capacity of this model is given by
the famous Shannon formula\footnote{%
Throughout this paper we use natural logarithms. In the context of quantum
channels \textquotedblleft capacity\textquotedblright\ will always mean the
\emph{classical capacity}.}
\begin{equation}
C=\frac{1}{2}\log \left( E+N\right) -\frac{1}{2}\log N=\frac{1}{2}\log
\left( 1+\frac{E}{N}\right) .  \label{shann}
\end{equation}

In the quantum analog of the signal+noise equation
\begin{equation*}
Y=X+Z
\end{equation*}%
one replaces the classical variables \ $X,\,Y,\,Z$ by a (multiple of) pair
of selfadjoint operators $q,p,$ satisfying the Heisenberg canonical
commutation relation (CCR) $[q,\,p]=i\hbar I,$ or, equivalently, by a single
operator $a=\frac{1}{\sqrt{2\hbar \omega }}\left( \omega q+ip\right) $ (with
Hermitean conjugate $a^{\dagger }=\frac{1}{\sqrt{2\hbar \omega }}\left(
\omega q-ip\right) $), satisfying the canonical commutation relation (CCR)
\begin{equation}
\lbrack a,\,a^{\dagger }]=I.  \label{ccra}
\end{equation}%
In applications $p$ and $q$ describe quantized quadratures of the harmonic
mode of frequency $\omega $,
\begin{equation}
q\omega\cos \omega t+ p\sin \omega t = \sqrt{\frac{\hbar \omega }{2}}
\left(a \mbox{e}^{-i\omega t}+a^{\dagger} \mbox{e}^{i\omega t}\right)\label{hm}
\end{equation}%
while $a,\,a^{\dagger }$ are quantizations of the complex amplitude and its
adjoint.

As distinct from the classical case, the quantum models should respect the
CCR i.e. arise as a part of a linear canonical transformation. Below we give
a list of such models which have additional property of gauge symmetry to be
explained later. In these models $a$ presents the quantum input signal, $b$
-- the quantum Gaussian noise variable and $a^{\prime }$ -- the quantum
output signal, all of them satisfying the CCR (\ref{ccra}).

\begin{enumerate}
\item Attenuator
\begin{equation*}
a^{\prime }=ka+\sqrt{1-k^{2}}b,\quad 0\leq k\leq 1 .
\end{equation*}%
$\quad $

\item Amplifier
\begin{equation*}
a^{\prime }=ka+\sqrt{k^{2}-1}b^{\dagger },\quad k\geq 1 .
\end{equation*}

\item Additive classical Gaussian noise%
\begin{equation*}
a^{\prime }=a+\eta ,
\end{equation*}%
where $\eta $ is the classical complex random variable having circular
Gaussian distribution.

\item Phase-invertive amplifier%
\begin{equation*}
a^{\prime }=ka^{\dagger }+\sqrt{k^{2}+1}b,\quad k\geq 0 .
\end{equation*}

\item Classical-quantum channel (state preparation)%
\begin{equation*}
a^{\prime }=x+b,
\end{equation*}%
where $x$ is the complex random variable representing classical signal at
the background quantum Gaussian noise $b$.

\item Quantum-classical channel (heterodyne measurement)%
\begin{equation*}
y=a+b^{\dagger },
\end{equation*}%
where $y$ is the complex random variable representing the classical output
of the measurement, $a$ is quantum signal, $b$ -- quantum Gaussian noise.
\end{enumerate}

All these equations have the form $Y=X+Z,$ where the noise $Z$ is described
by quantum or classical variable in Gaussian state\footnote{%
For detailed account of quantum Gaussian states see \cite{hol11}, \cite{hol13}.} with the
first two moments determined by%
\begin{equation}
\langle Z\rangle =0,\quad \langle Z^{\dagger }Z\rangle =N,\quad \langle
Z\,Z\rangle =0.  \label{2mom}
\end{equation}%
Thus in all the cases 1-6 the quantum AGWN model has the form (\ref{awgn})
where $Z_{k}$ are quantum or classical Gaussian i.i.d. noise variables
obeying (\ref{2mom}) and the signal sequence $X_{k}$ is subject to the
energy constraint
\begin{equation*}
n^{-1}\langle X_{1}^{\dag }X_{1}+\dots +X_{n}^{\dag }X_{n}\rangle \leq E.
\end{equation*}

A basic difference of the quantum signal variables is that one cannot simply
impose on them zero or other deterministic values; one should instead define
the \textit{state} describing these variables (in the classical case the
deterministic values are obtained from degenerate probability distributions).

This circumstance underlies a basic difficulty in finding the quantum analog
of the Shannon formula (\ref{shann}): finding the minimum of the output
entropy in the formula%
\begin{equation*}
C_{1}=\max_{\langle X^{\dagger }X\rangle \leq E}\,H(Y)-\min_{X}H(Y).\,
\end{equation*}%
Another problem is the proof of additivity of ``$n$-shot capacity'', $C_{n}=nC_{1}.$ When the signal $%
X $ is classical (case 5), the minimum $\min_{X}H(Y)=H(Z)$ is attained for $%
X\equiv 0.$ The resulting solution for the asymptotic capacity obtained in
\cite{hol98} is%
\begin{equation}
C=g\left( E+N\right) -g(N),  \label{ccq}
\end{equation}%
\newline
where
\begin{equation}
g(N)=(N+1)\log (N+1)-N\log N  \label{g}
\end{equation}%
is the function representing the entropy of quantum Gaussian state with the
moments (\ref{2mom}).

The capacity of heterodyne measurement (case 6) was obtained in \cite{hall}
by using a special \textquotedblleft
information-exclusion\textquotedblright\ method and is equal to
\begin{equation}
C=\log \left( E+N\right) -\log N.  \label{meas}
\end{equation}%
Alternatively, the minimal output entropy can be found using Lieb's solution
of Wehrl's conjecture \cite{hol} saying that the minimum is attained on the
coherent states.

In the cases 1-4 a similar \textquotedblleft Gaussian optimizers
conjecture\textquotedblright\ \cite{hw} was open for a dozen of years and
finally solved in \cite{ghg} (see Appendix 1). The resulting capacity
formula in the cases 1-3 has the same form as (\ref{ccq}), i.e.%
\begin{equation}
C=g\left( E+N\right) -g(N),\quad E=\langle X^{\dagger }X\rangle ,\quad
N=\langle Z^{\dagger }Z\rangle ,  \label{cgcov}
\end{equation}%
while in the case 4 it is%
\begin{equation}
C=g\left( E+N\right) -g(N+k^{2}).  \label{gcontra}
\end{equation}

All these solvable models possess symmetry under the gauge transformation $%
a\rightarrow a\,\mathrm{e}^{i\varphi },\,\varphi \in \mathbb{R}$. The
quantum channels 1-3, as well as \textquotedblleft hybrid\textquotedblright\
channels 5 (classical-quantum) and 6 (quantum-classical) are gauge-covariant
i.e. their output changes similarly to the input: $a^{\prime }\rightarrow
a^{\prime }\,\mathrm{e}^{i\varphi },$ while the channel 4 is
gauge-contravariant: $a^{\prime }\rightarrow a^{\prime }\,\mathrm{e}%
^{-i\varphi }.$ A complete classification of normal forms of single-mode
quantum Gaussian channels was given in \cite{hh}. In this classification the
cases 1-4 represent those normal forms which possess the gauge symmetry,
while 5,6 are the hybrid cases with this symmetry.

In the classical prototype of the gauge-covariant models $X,Y,Z$ are complex
Gaussian random variables having circular distribution and the capacity is
twice the Shannon expression (\ref{shann}) i.e. $C=\log \left( E+N\right)
-\log N.$

\section{The coding theorem for a broadband quantum channel}

In classical information theory the broadband channel can be treated by
reduction to parallel channels, i.e. by decomposing the Gaussian stochastic
process into independent one-dimensional harmonic modes (\ref{hm}). In
quantum theory such a decomposition plays an important additional role as a
tool for \textit{quantization} of the classical process. As a starting point
for the time-domain model of quantum noise we take the expression for
quantized electric field in a square box of size $L$\ (see, e.g. \cite{xhe})
\begin{equation*}
E(\mathbf{x},t)=\frac{i}{L^{3/2}}\sum_{k}\sqrt{\frac{\hbar \omega _{k}}{2}}%
a_{k}\mbox{e}^{i\mathbf{kx}}\mbox{e}^{-i\omega _{k}t}+\mathrm{h.c.}
\end{equation*}
where $a_{k}^{\dagger },a_{k}$ are the creation-annihilation operators of
independent bosonic modes satisfying the standard canonical commutation
relations\footnote{%
For simplicity we do not consider the polarization degree of freedom.}
\begin{equation}
[a_{j},a_{k}^{\dagger }]=\delta_{jk}I, \quad [a_{j},a_{k}]=0.
\end{equation}

Basing on this expression and redefining $a_{k}$, we consider the following
periodic operator-valued function as a model for observations on the time
interval $[0,T]$ at the spatial point $\mathbf{x=0}$:
\begin{equation}
\hat{Z}(t)=\sum_{k}\sqrt{\frac{\hbar \omega _{k}}{2T}}\left( a_{k}\mbox{e}%
^{-i\omega _{k}t}+a_{k}^{\dagger }\mbox{e}^{i\omega _{k}t}\right) ,\quad
t\in \lbrack 0,T],  \label{5-1}
\end{equation}%
\begin{equation*}
\omega _{k}=\frac{2\pi k}{T},\qquad k=1,2,...;\quad \Delta \omega =\frac{%
2\pi }{T},
\end{equation*}%
see \cite{hol98}. To avoid ultraviolet divergence, we introduce the cutoff
function $\bar{\omega}(T),T>0,$ with the properties: $\bar{\omega}(T)$ is
positive and monotonously increasing with $\lim_{T\rightarrow \infty }\bar{%
\omega}(T)=\infty ,$ and for each $T$ include in all summations over $k$
only the frequencies $\omega _{k}\in \lbrack 0,\bar{\omega}(T)].$ Then the
energy operator has the expression (as distinct from the narrowband
approximation):
\begin{equation*}
\int_{0}^{T}\hat{Z}(t)^{2}dt=\ \sum_{k}\hbar \omega _{k}\left(
a_{k}^{\dagger }a_{k}+{\frac{1}{2}}\right)
\end{equation*}

We modify the argument of \cite{hol98} related to classical-quantum channel
and generalize it to include the Gaussian gauge-covariant channels (cases
1-3). For a fixed $T$ the ``in-out'' equations of the channel $\Phi _{T}$ for the
collection of frequency modes are
\begin{equation}
a_{k,Y}=K(\omega _{k})a_{k,X}+\hat{n}_{k,Z},\quad 0\leq \omega _{k}\leq \bar{%
\omega}(T),  \label{chequation}
\end{equation}%
with the noise operators%
\begin{equation*}
\hat{n}_{k,Z}=\left\{
\begin{array}{ccc}
\sqrt{1-|K(\omega _{k})|^{2}}\,a_{k,Z}, & |K(\omega _{k})|\,<1 & \mathrm{%
attenuator} \\
\eta _{k,Z}, & |K(\omega _{k})|\,=1 & \mathrm{class.noise} \\
\sqrt{|K(\omega _{k})|^{2}-1}a_{k,Z}^{\dag }(\omega ) & |K(\omega _{k})|\,>1
& \mathrm{amplifier}%
\end{array}%
\right.
\end{equation*}%
satisfying the commutation relations%
\begin{equation*}
\lbrack \hat{n}_{k,Z}\,,\hat{n}_{l,Z}^{\dag }]=\delta _{kl}\left(
1-|K(\omega _{k})|^{2}\right) ,
\end{equation*}%
and described by a centered Gaussian state with the second moments%
\begin{equation*}
\quad \langle \hat{n}_{l,Z}^{\dag }\,\hat{n}_{k,Z}\rangle =\delta
_{kl}N(\omega _{k}),\quad \langle \hat{n}_{l,Z}\,\hat{n}_{k,Z}\rangle =0.
\end{equation*}%
Here $K(\omega ),N(\omega )$ are continuous functions, $N(\omega )\geq 0$ in
the domain $\omega \geq 0.$

Then $\Phi _{T}$ is Gaussian gauge-covariant channel in the Hilbert space $%
\mathcal{H}_{T}$ of the modes with frequencies $0\leq \omega _{k}\leq \bar{%
\omega}(T)$, whose action on the quantum states (density operators in $
\mathcal{H}_{T}$) is described in \cite{hh}, see also Ch. 12 of \cite{hol13}. We consider the family $\left\{ \Phi
_{T};T\rightarrow \infty \right\} $ as our model for the broadband channel.

\textbf{Definition}. For each $T>0$ a \textit{code} $(\Sigma ,M)$ is a
collection $\{\rho ^{j},M_{j};\,j=1\dots N\}$ where $\rho ^{j}$ are quantum
states in $\mathcal{H}_{T}$ satisfying the energy constraint\footnote{%
Notice that the vacuum energy $\frac{1}{2}\sum_{k}\hbar \omega _{k}$ is
explicitly excluded from the constraint to avoid the divergence when $%
T\rightarrow \infty .$}
\begin{equation}
\mathrm{Tr}\,\rho ^{j}\left( \sum_{k}\hbar \omega _{k}a_{k,X}^{\dag
}\,a_{k,X}\right) \leq ET,  \label{ec}
\end{equation}%
and $M$ is a POVM in $\mathcal{H}_{T}.$

We define the \textit{capacity} of the family $\left\{ \Phi
_{T};T\rightarrow \infty \right\} $ as the supremum of rates $R$ for which
the infimum of the average error probability
\begin{equation*}
{\bar{\lambda}}_{T}(\Sigma ,M)=\frac{1}{N}\sum_{j=1}^{N}(1-\mbox{Tr}\,\Phi
_{T}[\rho ^{j}]\,M_{j}).
\end{equation*}%
with respect to all codes of the size $N=\mbox{e}^{TR}$ tends to zero as $%
T\rightarrow \infty $.

\textbf{Theorem}. \textit{Let $N(\omega ),K(\omega )$ be continuous
functions, $0<|K(\omega )|\leq \kappa $, and } $\bar{\omega}(T)/T\rightarrow
0$ \textit{as} $T\rightarrow \infty $. \textit{The capacity of the family of
channels $\left\{ \Phi _{T};T\rightarrow \infty \right\} $ is equal to
\begin{equation}
C=\int_{0}^{\infty }(g(\tilde{N}_{\theta }(\omega ))-g(N(\omega )))_{+}\frac{%
d\omega }{2\pi },  \label{Cgc}
\end{equation}%
where%
\begin{equation*}
\tilde{N}_{\theta }(\omega )={\frac{1}{\mbox{e}^{\theta \hbar \omega
/\,|K(\omega )|^{2}}-1}},
\end{equation*}%
and $\theta $ is chosen such that
\begin{equation*}
\int_{0}^{\infty }(\hbar \omega \,/|K(\omega )|^{2})(\tilde{N}_{\theta
}(\omega )-N(\omega ))_{+}\frac{d\omega }{2\pi }=E.
\end{equation*}%
The capacity is upperbounded as
\begin{equation*}
C\leq \frac{\pi \kappa ^{2}}{6\hbar \,\theta }.
\end{equation*}%
}

The proof given in the Appendix 2 combines the solution of the quantum
Gaussian optimizer conjecture \cite{ghg} with the estimates from the proof
of the coding theorem for constrained infinite dimensional channel \cite%
{hol98}. The underlying mechanism is emergence of increasing number of
parallel independent channels in arbitrarily small neighbourhood of each
frequency. Similar proof applies to the classical capacities of time-domain
versions of gauge-contravariant channel (\ref{gcontra}) resulting in:
\begin{equation*}
C=\int_{0}^{\infty }(g(\tilde{N}_{\theta }(\omega ))-g(N(\omega )+|K(\omega
)|^{2}))_{+}\frac{d\omega }{2\pi }.
\end{equation*}%
For the case of quantum-classical channel (\ref{meas}) one has
\begin{equation*}
C=\int_{0}^{\infty }(\log N_{\theta }(\omega )-\log N(\omega ))_{+}\frac{%
d\omega }{2\pi },
\end{equation*}%
where $N_{\theta }(\omega )$ is given by (\ref{ntheta}) below.

For completeness we briefly recall here the case of classical-quantum
channel which was considered in \cite{hol98}. The channel equation in the
frequency domain is:%
\begin{equation*}
\Phi _{T}:\quad a_{k,Y}=x_{k}+a_{k,Z}.
\end{equation*}%
In this case it can be rewritten in the time domain as \textquotedblleft
classical signal + quantum noise\textquotedblright\ equation

\begin{equation*}
\hat{Y}(t)=X(t)+\hat{Z}(t),\quad t\in \lbrack -T/2,T/2],
\end{equation*}%
where the classical signal
\begin{equation*}
X(t)=\sum_{k}\sqrt{\frac{\hbar \omega _{k}}{2T}}\left( x_{k}\mbox{e}%
^{-i\omega _{k}t}+\bar{x}_{k}\mbox{e}^{i\omega _{k}t}\right) ,\quad x_{k}\in
\mathbb{C}.
\end{equation*}%
The mean power constraint on the signal
\begin{equation*}
\sum_{k}\hbar \omega _{k}|x_{k}|^{2}=\int_{0}^{T}X(t)^{2}dt\leq ET.
\end{equation*}%
Then with appropriate modification of Definition of the code, one obtains
the expression for the classical capacity
\begin{equation*}
C=\int_{0}^{\infty }(g(N_{\theta }(\omega ))-g(N(\omega )))_{+}\frac{d\omega
}{2\pi },
\end{equation*}%
\begin{equation}
N_{\theta }(\omega )={\frac{1}{\mbox{e}^{\theta \hbar \omega }-1}},
\label{ntheta}
\end{equation}%
and $\theta $ is chosen such that
\begin{equation*}
\int_{0}^{\infty }\hbar \omega (N_{\theta }(\omega )-N(\omega ))_{+}\frac{%
d\omega }{2\pi }=E.
\end{equation*}%
which coincides with the expression (\ref{Cgc}) for $K(\omega )\equiv 1.$

An example is the case of equilibrium quantum noise $N(\omega
)=N_{\theta _{P}}(\omega )\equiv (\mbox{e}^{\theta _{P}\hbar \omega
}-1)^{-1} $ with $\theta _{P}=\sqrt{\pi /12\hbar P}$ determined from
\begin{equation*}
\int_{0}^{\infty }\frac{\hbar \omega }{\mbox{e}^{\theta _{P}\hbar \omega }-1}%
\frac{d\omega }{2\pi }=P.
\end{equation*}%
Then%
\begin{equation*}
\int_{0}^{\infty }g\left( (\mbox{e}^{\theta _{P}\hbar \omega
}-1)^{-1}\right) =\frac{\pi }{6\hbar \,\theta _{P}}=\sqrt{\frac{\pi P}{%
3\hbar }}
\end{equation*}%
(see e.g. \cite{hol98} for detail of computation) and
\begin{equation*}
C=\sqrt{\frac{\pi (P+E)}{3\hbar }}-\sqrt{\frac{\pi P}{3\hbar }},
\end{equation*}%
which is similar to the capacity of the semiclassical broadband \emph{%
photonic channel} \cite{xle}, \cite{xcaves}.





\section{Discussion}

\subsection{The limiting broadband noise model}

\label{brb}

In the limit $T\rightarrow \infty $ of the periodic process (\ref{5-1})
converges in distribution to the quantum stationary Gaussian noise \cite%
{hol98}, \cite{xstat}
\begin{equation*}
\hat{Z}(t)=\int_{0}^{\infty }\sqrt{\frac{\hbar \omega }{2}}\left( d\hat{A}%
(\omega )\mbox{e}^{-i\omega t}+d\hat{A}(\omega )^{\dagger }\mbox{e}^{i\omega
t}\right) .
\end{equation*}%
Here $\hat{A}(\omega ),$ $\omega \geq 0,$ is the quantum Gaussian
independent increment process with the commutators
\begin{equation}
\lbrack d\hat{A}(\omega ),d\hat{A}(\omega ^{\prime })^{\dagger }]=\,\frac{1}{%
2\pi }\delta (\omega -\omega ^{\prime })d\omega \,d\omega ^{\prime },\quad
\lbrack d\hat{A}(\omega )\,,\,d\hat{A}(\omega ^{\prime })]=0,  \label{com}
\end{equation}%
zero mean, and the normally-ordered correlation
\begin{equation}
\langle d\hat{A}(\omega )^{\dagger }\,d\hat{A}(\omega ^{\prime })\rangle =\,%
\frac{1}{2\pi }\delta (\omega -\omega ^{\prime })N(\omega )d\omega \,d\omega
^{\prime }.  \label{cor}
\end{equation}%
This can be considered as an inhomogeneous generalization of the quantum
Brownian motion of Hudson-Parthasarathy \cite{hp}, albeit in the frequency
domain.

The noise commutator is causal
\begin{equation*}
\lbrack \hat{Z}(t),\hat{Z}(s)]=i\hbar /2\int_{0}^{\infty }\omega \sin \omega
(s-t)d\omega =i\hbar /2\delta ^{\prime }(t-s),
\end{equation*}%
and the noise symmetrized correlation function is%
\begin{equation*}
\alpha (t-s)\equiv \langle \hat{Z}(t)\circ \,\hat{Z}(s)\rangle =\beta (t-s)+
\frac{1}{2}j(t-s),
\end{equation*}%
where%
\begin{equation*}
\beta (t)=\hbar \int_{0}^{\infty }\omega N(\omega )\cos \omega t\,\frac{%
d\omega }{2\pi },
\end{equation*}%
\begin{equation*}
j(t)=\hbar \int_{0}^{\infty }\omega \cos \omega t\frac{d\omega }{2\pi }=-%
\frac{\hbar }{2\pi }t^{-2}.
\end{equation*}%
so that the vacuum symmetrized correlation function $\frac{1}{2}j(t-s)$.

One can then introduce the gauge-covariant channels in the frequency domain
by the equation%
\begin{equation}
d\hat{A}_{Y}(\omega )=K(\omega )d\hat{A}_{X}(\omega )+d\hat{A}_{Z}(\omega ),
\label{frch}
\end{equation}%
where the appropriately modified Gaussian noise $\hat{Z}(t)$ satisfies (cf.
\cite{bs})
\begin{eqnarray*}
\lbrack d\hat{A}_{Z}(\omega ),\,d\hat{A}_{Z}^{\dag }(\omega ^{\prime })]
&=&\,\frac{1}{2\pi }\delta (\omega -\omega ^{\prime })(1-|K(\omega
)|^{2})d\omega \,d\omega ^{\prime },\quad \\
\langle d\hat{A}_{Z}^{\dag }(\omega )\,\,d\hat{A}_{Z}(\omega ^{\prime
})\rangle &=&\,\frac{1}{2\pi }\delta (\omega -\omega ^{\prime })N(\omega
)d\omega \,d\omega ^{\prime }.
\end{eqnarray*}%
In the time domain, asymptotically (as $T\rightarrow \infty $)%
\begin{equation}
\hat{Y}(t)\approx (KX\mathbf{)}(t)+\hat{Z}(t),  \label{appr}
\end{equation}%
with nonanticipating real-valued filter
\begin{equation*}
(K\hat{X})(t)=\int^{t}\hat{X}(s)k(t-s)ds,\quad K(\omega )=\int_{0}^{\infty
}k(t)\mbox{e}^{i\omega t}dt=\overline{K(-\omega )}.
\end{equation*}%
If $K$ is instantaneous or has finite memory, then (\ref{appr}) becomes
equality.

The noise is \emph{generalized} quantum (operator-valued) Gaussian process, $%
R(f)=\int_{-\infty }^{\infty }\hat{Z}(t)\,f(t)dt,$ where $\,f$ runs over an
appropriate space of test functions. The mathematical construction which
gives to it a rigorous meaning is based on quasi-free representations of the
C*-algebra $\mathfrak{A}(\mathcal{H},\Delta )$ of CCR \cite{man} over the
symplectic space $\mathcal{H}=\mathcal{K}(\mathbb{R})$ of real-valued
infinite differentiable functions with compact support, with the
skew-symmetric form $\Delta $, and the vacuum inner product $j$, given by $%
(\hbar =2)$
\begin{eqnarray*}
\Delta (f,g) &=&\int_{-\infty }^{\infty }f(t)\frac{d}{dt}g(t)\,dt=\pi ^{-1}%
\mathrm{Im}\int_{0}^{\infty }\omega \,\overline{\tilde{f}(\omega )}\tilde{g}%
(\omega )\,d\omega , \\
j(f,g) &=&\pi ^{-1}\mathrm{Re}\int_{0}^{\infty }\omega \,\overline{\tilde{f}%
(\omega )}\tilde{g}(\omega )\,d\omega
\end{eqnarray*}%
\begin{equation*}
=\pi ^{-1}\int_{-\infty }^{\infty }g(t)\int_{-\infty }^{\infty }\frac{%
2f(t)-f(t-s)-f(t+s)}{s^{2}}ds\,dt
\end{equation*}%
\begin{equation}
=(2\pi )^{-1}\int_{-\infty }^{\infty }\int_{-\infty }^{\infty
}(g(t)-g(t-s))(f(t)-f(t-s))s^{-2}ds\,dt,  \label{slob}
\end{equation}%
see n.7.1 of \cite{xstat}. The operator of complex structure $J$ is
multiplication by $i\,\mathrm{sgn}(\omega )$ in the frequency domain or the
Hilbert transform in the time domain
\begin{equation*}
(Jz)(t)=\frac{1}{\pi }\mathrm{V.P.}\int_{-\infty }^{\infty }\frac{z(s)}{t-s}%
\,ds
\end{equation*}

Here $H=\overline{K_{\infty }(\mathbb{R})}$, the completion is with respect
to Slobodeckij (semi)norm, corresponding to the inner product (\ref{slob})
\cite{sl}. Thus the relevant space is Sobolev-Slobodeckij space $H^{1/2}(%
\mathbb{R})$ of half-differentiable functions.

A natural conjecture would be that the asymptotic (as $T\rightarrow \infty $%
) capacity of the channel (\ref{appr}) over observations in the subspace $%
\mathcal{H}_{T}=\mathcal{K}([0,T])$ of test functions with support in $[0,T]$
is given by the expression (\ref{Cgc}) from the coding theorem above. Such a
proof would be free from a simplification inherent in our model due to the
assumed independence of the modes $a_{k}$ for each $T$.

However an attempt to adapt the classical proof \cite{xgal} meets obstacles
arising from the additional symplectic structure and the fact that the
observation subspace $\mathcal{H}_{T}$ is \textit{not invariant} under the
complex structure $J.$ Such kind of problems do not arise in the
\textquotedblleft narrowband\textquotedblright\ approximations of the type
considered in \cite{bs} where the Planck vacuum spectrum is replaced by the
flat one. A discussion of such a noise model is given is the next section.

\subsection{Bandpass noise model}

We mentioned that in the classical prototype of the gauge-covariant models
1-6, $X,Y,Z$ are complex Gaussian random variables having circular
distribution. This suggests to consider the following quantum noise model%
\begin{equation*}
\mathbf{\hat{Z}}(t)=\int_{-\infty }^{\infty }\mbox{e}^{i\omega t}d\hat{A}%
(\omega ),
\end{equation*}%
where $\hat{A}(\omega ),$ $\omega \in \mathbb{R},$ the quantum Gaussian
independent increment process with the commutators (\ref{com}) and
correlation (\ref{cor}), but on the whole real line, with the spectral
density $N(\omega )\geq 0,$ $\omega \in \mathbb{R}$. The noise has causal
commutator
\begin{equation}
\lbrack \mathbf{\hat{Z}}(t),\,\,\mathbf{\hat{Z}}^{\dag }(s)]=\,\frac{1}{2\pi
}\int_{-\infty }^{\infty }\mbox{e}^{i\omega (t-s)}\,\,d\omega =\delta \left(
t-s\right) .  \label{cdelta}
\end{equation}%
The noise is thus generalized operator-valued process.

Then the \emph{normally ordered} correlation function of the noise is%
\begin{equation*}
\boldsymbol{ \beta (t-s)}=\langle \mathbf{\hat{Z}}\,^{\dag }(s)\,\mathbf{%
\hat{Z}}(t)\,\rangle =\int_{-\infty }^{\infty }\mbox{e}^{i\omega
(t-s)}N(\omega )\,\,\frac{d\omega }{2\pi }.
\end{equation*}

Introducing the models 1-3 of gauge-covariant channels with filtered signal
in the frequency domain by the relation (\ref{frch}) for all real $\omega $
we arrive to the \textquotedblleft time-invariant\textquotedblright\ model
considered by Bardhan and Shapiro in \cite{bs} for a special noise spectral
density ${N(\omega ).}$

A formal solution for the asymptotic capacity of such model is given by the
relation similar to (\ref{Cgc}) but without $\hbar \omega ,$ namely:
\begin{equation*}
C=\int \left( g(\bar{N}_{\theta }(\omega ))-g(N(\omega ))\right) _{+}\frac{%
d\omega }{2\pi },
\end{equation*}%
where%
\begin{equation*}
\bar{N}_{\theta }(\omega )={\frac{1}{\mbox{e}^{\theta \,|K(\omega )|^{-2}}-1}%
}
\end{equation*}%
and $\theta $ is determined from%
\begin{equation*}
\int \,|K(\omega )|^{-2}(\bar{N}_{\theta }(\omega )-N(\omega ))_{+}\,\frac{%
d\omega }{2\pi }=E.
\end{equation*}

Notice that the capacity may be infinite unless $\,|K(\omega )|$ decreases
fast enough as $\omega \rightarrow \infty $ (see Appendix 3). We anticipate
that a general proof for this model can be given along the same lines as in
the classical case \cite{xgal} due to the special form of the commutator (%
\ref{cdelta}) which agrees with the simple complex structure of
multiplication by $i$. Indeed, the relevant symplectic space is $\mathcal{H}%
=L_{\mathbb{C}}^{2}(\mathbb{R}_{+}),$ considered as \textit{real} vector
space with the skew-symmetric form and the vacuum inner product,
correspondingly,
\begin{eqnarray*}
\Delta (f,g) &=&\mathrm{Im}\,\int_{0}^{\infty }\overline{f(t)}%
\,g(t)\,dt=\left( 2\pi \right) ^{-1}\mathrm{Im}\int_{-\infty }^{\infty }%
\overline{\tilde{f}(\omega )}\tilde{g}(\omega )d\omega \\
j(f,g) &=&\mathrm{Re}\,\int_{0}^{\infty }\overline{f(t)}\,g(t)\,dt=\left(
2\pi \right) ^{-1}\mathrm{Re}\int_{-\infty }^{\infty }\overline{\tilde{f}%
(\omega )}\tilde{g}(\omega )d\omega ,
\end{eqnarray*}%
and $J$ is just multiplication by $i$. The subspace $\mathcal{H}_{T}=L_{%
\mathbb{C}}^{2}([0,T])$ is \textit{invariant} under $J.$ Therefore the
argument can be essentially a complexified version using the spectral theory
for operators with Hermitian (rather than real symmetric) continuous kernels
on $[0,T]$ for the decomposition into normal modes, and the corresponding
Kahrunen-Loewe expansion.

As argued in \cite{bs}, such an approach is suitable for narrowband
channels. A question that arises naturally is a derivation of this bandpass
model from the broadband model of Sec. \ref{brb} under certain precise
limiting conditions.

\subsection{The symplectic eigenvalue problem}

In the finite-dimensional case the normal mode decomposition is closely
related to finding symplectic eigenvalues of the correlation matrix $\alpha$%
, which can be defined as numbers $\lambda$ satisfying
\begin{equation*}
\lbrack \alpha -i\lambda \Delta ]f=0
\end{equation*}
for some $f\neq 0$. The continuous-time analog of this is the integral
equation
\begin{equation}  \label{sep}
\lbrack \alpha _{T}-i\lambda \Delta _{T}]f=0,
\end{equation}
where $\alpha _{T}$ is the integral operator with the symmetric kernel $%
\alpha (t-s) =\beta (t-s)+ \frac{1}{2}j(t-s)$ on $[0, T]$, and $f\neq 0$
belongs to certain completion of the space $K_T$.

Next the problem arises to show that for $T\rightarrow \infty $ the
symplectic eigenvalues tend to the continuous spectral distribution.

The difference between the two models appears here most apparently:

\underline{Bandpass model}: by complexification, (\ref{sep}) reduces to
ordinary eigenvalue problem for the continuous hermitean kernel $%
\boldsymbol{\beta }(t-s)=\int_{-\infty }^{\infty }\mathrm{e}^{i\omega
(t-s)}N(\omega )\,\frac{d\omega }{2\pi }$:
\begin{equation*}
\int_{0}^{T}\boldsymbol{\beta }(t-s)f(s)ds=(\lambda -1/2)f(t),\quad t\in
\lbrack 0,T];
\end{equation*}%
The limit $T\rightarrow \infty $ can be treated as in the Kac-Merdock-Sz\"{o}%
ge theorem.

\underline{Broadband model}: the symplectic eigenvalue equation (\ref{sep})
takes the form
\begin{equation*}
\int_{0}^{T}\beta (t-s)f(s)ds+\frac{1}{2\pi }\int_{0}^{T}\frac{%
2f(t)-f(t-s)-f(t+s)}{s^{2}}ds\,=i\lambda f^{\prime }(t),\quad t\in \lbrack
0,T].
\end{equation*}
The study of such an equation is a subject of a future work.

\section*{Appendix 1}

We denote by $S(\rho )=-\mbox{Tr}\rho \log \rho $ the von Neumann entropy.
Let $H$ be a positive selfadjoint operator, representing the energy. The
constrained $\chi $-capacity of a channel $\Phi $ can be expressed as \cite%
{shir}
\begin{equation}
C_{\chi }(\Phi ;H,E)=\sup_{\mathrm{tr\,}\bar{\rho}_{\pi }\,H\leq E}\left\{
S(\Phi \lbrack \mathrm{\,}\bar{\rho}_{\pi }])-\int S(\Phi \lbrack \rho ])\pi
(d\rho )\right\} ,  \label{oneshotunassist}
\end{equation}%
where the maximization is performed over the set of input ensembles $\pi $
(probability distributions on the set of quantum states (density operators) $%
\rho $) satisfying the constraint $\mathrm{tr\,}\bar{\rho}_{\pi }\,H\leq E$
where $\bar{\rho}_{\pi }=\int \rho \,\pi (d\rho )$ is the average state of
the ensemble.

Let $\Phi $ be an $s-$mode Gaussian gauge-covariant channel defined by the
matrix parameters $\mathbf{K},\boldsymbol{\mu}$ as in \cite{ghg}, and $%
H=\sum \epsilon _{kl}a_{k}^{\dag }a_{l}$ a quadratic gauge-invariant
Hamiltonian with Hermitean energy matrix $\boldsymbol{\nu }=[\epsilon _{kl}]$%
. By the solution of the quantum Gaussian optimizers conjecture \cite{ghg},
the quantity $C_{\chi }(\Phi ;H,E)$ is given by

\begin{eqnarray}
&&C_{\chi }(\Phi ;H,E)  \label{maxnu} \\
&=&\max_{\boldsymbol{\nu }:\,\mathrm{tr}\boldsymbol{\nu \epsilon }\leq E}\,%
\mathrm{\ tr\,}g(\mathbf{K}^{\ast }\boldsymbol{\nu}\mathbf{K}+%
\boldsymbol{\mu}+\left( \mathbf{K}^{\ast }\mathbf{K}-\mathbf{I}_{B}\right)
/2)-\mathrm{tr}g(\boldsymbol{\mu}+\left( \mathbf{K}^{\ast }\mathbf{K}-%
\mathbf{I}_{B}\right) /2).  \notag
\end{eqnarray}%
The optimal ensemble $\pi _{\ast }$ which attains the supremum in (\ref%
{oneshotunassist}) consists of coherent states $\rho _{\mathbf{x}},\,\mathbf{%
x}\in \mathbb{C}^{s}$ distributed with gauge-invariant Gaussian probability
distribution $\pi _{\boldsymbol{\nu }}(d^{2s}x)$ on $\mathbb{C}^{s}$ having
zero mean and the correlation matrix $\boldsymbol{\nu }$ which solves the
maximization problem in (\ref{maxnu}). Here $\rho _{0}$ is the vacuum
density operator.

Next, let $\Phi _{k};\,k=1,\dots ,n,$ be gauge-covariant Gaussian channels, $%
H_{k}$ the quadratic Hamiltonians. Put $\Phi =\Phi _{1}\otimes \dots \otimes
\Phi _{n}$ and$\ H=H_{1}\otimes I\dots \otimes I+\dots +I\otimes \dots
\otimes I\otimes H_{n}.$ Then the parameters $\mathbf{K},\boldsymbol{\mu}$
of $\Phi $ have block-diagonal form. From (\ref{maxnu}) it follows that the
maximizing $\boldsymbol{\nu }$ also have the block-diagonal form and hence
the additivity property follows

\begin{equation}
C_{\chi }(\Phi ;H,E)=\max_{E_{1}+\dots +E_{n}\leq E}\left[ C_{\chi }(\Phi
_{1},H_{1},E_{1})+\dots +C_{\chi }(\Phi _{n},H_{n},E_{n})\right] .
\label{clc2}
\end{equation}

Consider the special case of independent uncorrelated noise modes, where $%
\Phi _{k}$ is the one-mode channel in the canonical form given by the
equation (\ref{chequation}) with the energy operator $H_{k}=\hbar \omega
_{k}a_{k,X}^{\dag }\,a_{k,X}.$ In this case $E_{k}=\hbar \omega _{k}m_{k}$
and
\begin{equation*}
C_{\chi }(\Phi _{k},H_{k},E_{k})=g(|K(\omega _{k})|^{2}m_{k}+N(\omega
_{k}))-g(N(\omega _{k}))
\end{equation*}%
so that (\ref{clc2}) reduces to

\begin{equation}
C_{\chi }(\Phi ;H,E)=\max \sum_{k}\left[ g(|K(\omega
_{k})|^{2}m_{k}+N(\omega _{k}))-g(N(\omega _{k}))\right] ,
\end{equation}%
where the maximum is taken over the set
\begin{equation*}
m_{k}\geq 0,\qquad \sum_{k}\hbar \omega _{k}m_{k}\leq E.
\end{equation*}

Using the Kuhn-Tucker condition for the optimization of the concave
function, which has the form%
\begin{equation*}
|K(\omega _{k})|^{2}g^{\prime }(|K(\omega _{k})|^{2}m_{k}+N(\omega
_{k}))-\theta \hbar \omega _{k}\leq 0
\end{equation*}%
with the equality if and only if$m_{k}>0,$ and taking into account that $%
g^{\prime }(x)=\log (1+x^{-1}),$ one finds the \textquotedblleft quantum
water-filling\textquotedblright\ solution (cf. \cite{hol98}, \cite{bs})
\begin{eqnarray}
C_{\chi }(\Phi ;H,E) &=&\sum_{k}[g(|K(\omega _{k})|^{2}m_{k}^{\ast
}+N(\omega _{k}))-g(N(\omega _{k}))]  \label{wfs} \\
&=&\sum_{k}\left[ g\left( {\frac{1}{\mbox{e}^{\theta \hbar \omega
_{k}/|K(\omega _{k})|^{2}}-1}}\right) -g(N(\omega _{k}))\right] _{+},  \notag
\end{eqnarray}%
where
\begin{equation*}
m_{k}^{\ast }=|K(\omega _{k})|^{-2}\left( {\frac{1}{\mbox{e}^{\theta \hbar
\omega _{k}/|K(\omega _{k})|^{2}}-1}}-N(\omega _{k})\right) _{+},\quad
\end{equation*}%
and $\theta $ is chosen in such a way that
\begin{equation*}
\sum_{k}\hbar \omega _{k}m_{k}^{\ast }=E.
\end{equation*}

\section*{Appendix 2. Proof of the Coding Theorem}

We first prove the weak converse:%
\begin{equation}
\inf_{\Sigma ,M}{\bar{\lambda}}_{T}(\Sigma ,M)\not\rightarrow 0\quad \mathrm{%
for}\quad R>C.  \label{wc}
\end{equation}%
From the classical Fano inequality and the quantum bound for classical
information 
\begin{equation}
TR\cdot (1-\inf_{\Sigma ,M}{\bar{\lambda}}_{T}(\Sigma ,M))\leq C_{\chi ,T}+1,
\label{5-12}
\end{equation}%
where $C_{\chi ,T}$ is the constrained $\chi -$capacity (\ref%
{oneshotunassist}) of the channel $\Phi _{T}$ in $\mathcal{H}_{T}$, so that $%
C_{\chi ,T}=C_{\chi }(\Phi _{T};H,ET)$ given by (\ref{wfs}).

Taking into account that
\begin{equation}
\Delta \omega =\frac{2\pi }{T},  \label{5-14}
\end{equation}%
the energy constraint can be rewritten as
\begin{equation*}
\sum_{k}\hbar \omega _{k}m_{k}{\frac{\Delta \omega }{2\pi }}\leq E.
\end{equation*}%
The \textquotedblleft quantum water-filling\textquotedblright\ solution is
then
\begin{equation}
\frac{C_{\chi ,T}}{T}=\sum_{k}[g(|K(\omega _{k})|^{2}m_{k}^{\ast }+N(\omega
_{k}))-g(N(\omega _{k}))]\frac{\Delta \omega }{2\pi },  \label{5-27}
\end{equation}%
where
\begin{equation}
m_{k}^{\ast }=|K(\omega _{k})|^{-2}\left( {\frac{1}{\mbox{e}^{\theta
_{T}f(\omega _{k})}-1}}-N(\omega _{k})\right) _{+},  \label{mk*}
\end{equation}%
and $\quad $%
\begin{equation}
f(\omega )=\hbar \omega /|K(\omega )|^{2}\geq \frac{\hbar \omega }{\kappa
^{2}},  \label{b}
\end{equation}%
while $\theta _{T}$ is chosen in such a way that
\begin{equation}
\sum_{k}\hbar \omega _{k}m_{k}^{\ast }{\frac{\Delta \omega }{2\pi }}=E.
\label{5-28}
\end{equation}%
By considering the piecewise constant functions
\begin{equation*}
N_{T}(\omega )=N(\omega _{k}),\quad K_{T}(\omega )=K(\omega _{k}),\quad
f_{T}(\omega )=f(\omega _{k}),\quad \omega _{k-1}<\omega \leq \omega
_{k},\quad k=0,1,\dots
\end{equation*}%
and%
\begin{equation*}
m_{T}(\omega )=m_{k}^{\ast }\qquad \omega _{k-1}<\omega \leq \omega _{k}\leq
\bar{\omega}(T);\quad m_{T}(\omega )=0,\qquad \omega >\bar{\omega}(T),
\end{equation*}%
we can write the right hand side of (\ref{5-27}) as
\begin{equation*}
\int_{0}^{\infty }[g(N_{T}(\omega )+|K_{T}(\omega )|^{2}m_{T}(\omega
))-g(N_{T}(\omega ))]\frac{d\omega }{2\pi }
\end{equation*}%
\begin{equation*}
=\int_{0}^{\infty }[g(N(\omega )+|K(\omega )|^{2}m_{T}(\omega ))-g(N(\omega
))]\frac{d\omega }{2\pi }
\end{equation*}%
\begin{equation}
+\int_{0}^{\infty }[g(N_{T}(\omega )+|K_{T}(\omega )|^{2}m_{T}(\omega
))-g(N(\omega )+|K(\omega )|^{2}m_{T}(\omega ))+g(N(\omega )-g(N_{T}(\omega
))]\frac{d\omega }{2\pi }.  \label{2ndterm}
\end{equation}%
Taking into account that
\begin{equation}
\int_{0}^{\infty }\hbar \omega m_{T}(\omega )\frac{d\omega }{2\pi }\leq
\sum_{k}\hbar \omega _{k}m_{k}^{\ast }\frac{\Delta \omega }{2\pi }=E,
\label{star}
\end{equation}%
we see that the first term in the right hand side of (\ref{2ndterm}) is less
than or equals to
\begin{equation*}
\max_{m\in \mathcal{M}}\int_{0}^{\infty }[g(N(\omega )+|K(\omega
)|^{2}m(\omega ))-g(N(\omega ))]\frac{d\omega }{2\pi },
\end{equation*}%
where
\begin{equation*}
\mathcal{M}=\{m(\cdot ):m(\omega )\geq 0,\quad \int_{0}^{\infty }\hbar
\omega m(\omega )\frac{d\omega }{2\pi }\leq E\}.
\end{equation*}%
Similarly to (\ref{mk*}), the solution is given by the function
\begin{equation}
m^{\ast }(\omega )=|K(\omega )|^{-2}\left( \tilde{N}_{\theta }(\omega
)-N(\omega )\right) _{+},  \label{5-15}
\end{equation}%
where $\theta $ is determined from%
\begin{equation*}
\int_{0}^{\infty }\hbar \omega m^{\ast }(\omega )\frac{d\omega }{2\pi }=E,
\end{equation*}%
in other words, the first term is less than or equals to $C.$

If we show that the second term in (\ref{2ndterm}) tends to zero then we
will have
\begin{equation}
{\limsup }_{T\rightarrow \infty }\frac{C_{\chi ,T}}{T}\leq C  \label{cbound}
\end{equation}%
and therefore from (\ref{5-12})
\begin{equation*}
(1-{\lim \inf }_{T\rightarrow \infty }\inf_{\Sigma ,M}{\bar{\lambda}}%
_{T}(\Sigma ,M))\leq C/R,
\end{equation*}
hence the weak converse (\ref{wc}).

We shall show it by using the Lebesgue dominated convergence theorem. Since $%
N(\omega )$ is continuous, $N_{T}(\omega )\rightarrow N(\omega )$ and $%
g(N_{T}(\omega ))\rightarrow g(N(\omega ))$ pointwise. Next we observe that $%
\theta _{T}$ is separated from $0$ as $T\rightarrow \infty $, that is $%
\theta _{T}\geq \theta _{\infty }>0$. Indeed, assume that $\theta
_{T}\downarrow 0$ for some sequence $T\rightarrow \infty $, then the
sequence of continuous functions
\begin{equation}
m_{T}(\omega )=|K_{T}(\omega )|^{-2}\left( {\frac{1}{\mbox{e}^{\theta
_{T}f_{T}(\omega )}-1}}-N_{T}(\omega )\right) _{+}  \label{a}
\end{equation}%
converges to $\infty $ uniformly in every interval $0<{\underline{\omega }}%
\leq \omega \leq {\bar{\omega}}<\infty $, which contradicts to the condition
(\ref{star}). It follows that for any fixed $\omega >0$ the quantity
\begin{equation*}
N_{T}(\omega )+|K_{T}(\omega )|^{2}m_{T}(\omega )=\max \left( {\frac{1}{%
\mbox{e}^{\theta _{T}f_{T}(\omega )}-1}},N_{T}(\omega )\right)
\end{equation*}%
is bounded as $T\rightarrow \infty $. Since $g(x)$ is uniformly continuous
on any bounded interval, it follows that
\begin{equation*}
g(N_{T}(\omega )+|K_{T}(\omega )|^{2}m_{T}(\omega ))-g(N(\omega )+|K(\omega
)|^{2}m_{T}(\omega ))\rightarrow 0
\end{equation*}%
pointwise.

Let us show that the integrand is dominated by an integrable function.
Taking into account that $g^{\prime }(x)$ $\geq 0$ and $g^{\prime \prime
}(x)\leq 0$ for $x\geq 0$, we deduce that $0\leq g(x+y)-g(x)\leq g(y)$ for $%
x,y\geq 0$. Therefore the integrand is dominated from above by the function $%
g(|K_{T}(\omega )|^{2}m_{T}(\omega ))$ and from below by the function $%
-g(|K(\omega )|^{2}m_{T}(\omega ))$. But from (\ref{a}), (\ref{b})
\begin{equation}
|K_{T}(\omega )|^{2}m_{T}(\omega )\leq {\frac{1}{\mbox{e}^{\theta
_{T}f_{T}(\omega )}-1}}\leq {\frac{1}{\mbox{e}^{\theta _{\infty
}f_{T}(\omega )}-1}}\leq {\frac{1}{\mbox{e}^{c_{\infty }\omega }-1}}
\label{ineq}
\end{equation}%
with $c_{\infty }=\theta _{\infty }\hbar /\kappa ^{2}>0.$ Thus
\begin{equation*}
g(|K_{T}(\omega )|^{2}m_{T}(\omega ))\leq g\left( {\frac{1}{\mbox{e}%
^{c_{\infty }\omega }-1}}\right) ={\frac{c_{\infty }\omega }{\mbox{e}%
^{c_{\infty }\omega }-1}}-\log (1-\mbox{e}^{-c_{\infty }\omega }),
\end{equation*}%
which is positive integrable function. There is also a similar estimate from
below. Thus (\ref{cbound}) follows establishing the weak converse. The last
inequality also implies that integrand in (\ref{Cgc}) is upperbounded by
integrable function proving finiteness of the capacity, namely
\begin{equation*}
C\leq \int_{0}^{\infty }g\left( {\frac{1}{\mbox{e}^{c_{\infty }\omega }-1}}%
\right) \frac{d\omega }{2\pi }=\frac{\pi }{6c_{\infty }}=\frac{\pi \kappa
^{2}}{6\hbar \,\theta _{\infty }}.
\end{equation*}

We now proceed to prove the direct statement of the coding theorem: for
appropriately chosen codes the average error probability tends to zero when $%
T\rightarrow \infty $ and $R<C$. Let us introduce some notations.

Denote $\rho _{x_{k}}=|x_{k}\rangle \langle x_{k}|;\,x_{k}\in \mathbb{C}$,
the coherent state for the $k$-th mode, and $\rho _{x}=\otimes _{k}\rho
_{x_{k}};\,x=\left\{ x_{k}\right\} $ the coherent state for the collection
of all modes with $\omega _{k}\in \lbrack 0,\bar{\omega}(T)],$ so that the
number of the modes is equal to $s_{T}=\frac{\bar{\omega}(T)T}{2\pi }.$ In
particular $\rho _{0}$ is the vacuum state. We consider the codebooks of the
form $\Sigma =\left\{ \rho _{x^{1}},\dots ,\rho _{x^{N}}\right\} $ and
denote $\rho _{j}^{\prime }=\Phi _{T}[\rho _{x^{j}}].$ It is Gaussian
diagonal state with mean $x^{j}=\left\{ x_{k}^{j}\right\} $ and photon
numbers $\left\{ N(\omega _{k})\right\} .$

Let $\pi _{\ast }(d^{2s_{T}}x)$ be the Gaussian probability distribution
\begin{equation}
\pi _{\ast }(d^{2s_{T}}x)=\mbox{exp}\left( -\sum_{k}{\frac{|x_{k}|^{2}}{%
m_{k}^{\ast }}}\right) \prod_{k}d^{2}x_{k},  \label{4-7}
\end{equation}%
where $m_{k}^{\ast }$ are given by (\ref{mk*}). (If $m_{k}^{\ast }=0$, we
have in mind in (\ref{4-7}) the Gaussian distribution degenerated at $0$.) $%
\ \pi _{\ast }$ is the optimal distribution on the coherent states on which $%
C_{\chi ,T}$ is achieved in (\ref{5-27}).

Denote $\bar{\rho}_{\ast }^{\prime }=\Phi _{T}[\bar{\rho}_{\pi }].$ It is
Gaussian diagonal state with mean $0$ and photon numbers%
\begin{equation*}
N_{k}^{\prime }=|K(\omega _{k})|^{2}m_{k}^{\ast }+N(\omega _{k})=\max
\left\{ {\frac{1}{\mbox{e}^{\theta _{T}f(\omega _{k})}-1}},N(\omega
_{k})\right\} .
\end{equation*}%
Define the suboptimal decoding $M=\left\{ M^{1},\dots ,M^{N}\right\} $
similarly to Eq. (44) in the proof of the coding theorem in \cite{hol98}:%
\begin{equation}
M^{j}=(\sum_{l=1}^{N}PP^{l}P)^{-\frac{1}{2}}\left( PP^{j}P\right)
(\sum_{l=1}^{N}PP^{l}P)^{-\frac{1}{2}},  \label{gs1}
\end{equation}%
where, however, $P$ is the spectral projection of $\bar{\rho}_{\ast
}^{\prime }$ corresponding to the eigenvalues in the range $(\mbox{e}^{-[H({%
\bar{\rho}}_{\ast }^{\prime })+\delta T]},\mbox{e}^{-[H({\bar{\rho}}_{\ast
}^{\prime })-\delta T]})$, and $P^{j}$ is the spectral projection of $\rho
_{j}^{\prime }$ corresponding to the eigenvalues in the range $(\mbox{e}%
^{-[H(\rho _{0})+\delta T]},\mbox{e}^{-[H(\rho _{0})-\delta T]})$. Since $%
\rho _{j}^{\prime }$ are all unitarily equivalent to $\rho _{0}^{\prime }$,
then $H(\rho _{j}^{\prime })=H(\rho _{0}^{\prime }),$ where $\rho
_{0}^{\prime }=\Phi _{T}[\rho _{0}]$ is Gaussian diagonal state with mean $0$
and photon numbers $\left\{ N(\omega _{k})\right\} .$

Applying the basic inequality Eq. (50) from \cite{hol98} with the word
length $n=1$ and with $\delta $ replaced by $\delta T$, we have
\begin{equation}
\inf_{M}{\bar{\lambda}}(\Sigma ,M)\leq  \label{5-17}
\end{equation}%
\begin{equation*}
\leq \frac{1}{N}\sum_{j=1}^{N}\left\{ 3\mbox{Tr}\rho _{j}^{\prime }(I-P)+%
\mbox{Tr}\rho _{j}^{\prime }(I-P_{x^{j}})+\sum_{l\neq j}\mbox{Tr}P\rho
_{j}^{\prime }PP^{l}\right\} ,
\end{equation*}%
Since $\rho _{j}$ are unitary equivalent to $\rho _{0}$, then the middle
term in (\ref{5-17}) is simply%
\begin{equation}
\mbox{Tr}\rho _{0}^{\prime }(I-P_{0}),  \label{5-18}
\end{equation}%
which is similar to
\begin{equation}
\mbox{Tr}{\bar{\rho}}_{\ast }^{\prime }(I-P).  \label{5-19}
\end{equation}

We wish to estimate the terms (\ref{5-18}), (\ref{5-19}) for the Gaussian
density operators $\rho _{0}^{\prime },\,${$\bar{\rho}^{\prime }$}$_{\pi }$.
For definiteness let us take (\ref{5-18}). We have
\begin{equation}
\mbox{Tr}\rho _{0}^{\prime }(I-P_{0})=\mathsf{Pr}\left\{ |-\log \lambda
_{(\cdot )}-H(\rho _{0})|\geq \delta T\right\} ,  \label{5-20}
\end{equation}%
where $\mathsf{Pr}$ is the distribution of eigenvalues $\lambda _{(\cdot )}$
of $\rho _{0}^{\prime }$. By Chebyshev inequality, this is less or equal to $%
\mathsf{D}(\log \lambda _{(\cdot )})/\delta ^{2}T^{2}$. Now $\mathsf{D}(\log
\lambda _{(\cdot )})=\sum_{k}\mathsf{D}_{k}(\log \lambda _{(\cdot )})$,
where $\mathsf{D}_{k}$ is the variance of $\log \lambda _{(\cdot )}$ for the
$k$-th mode. The eigenvalues of the Gaussian density operator $\rho
_{k}^{\prime }(0)$ are
\begin{equation*}
\lambda _{n}^{k}=\frac{N(\omega _{k})^{n}}{(N(\omega _{k})+1)^{n+1}};\qquad
n=0,1,...,
\end{equation*}%
hence
\begin{equation}
\mathsf{D}_{k}(\log \lambda _{(\cdot )})=\sum_{n=0}^{\infty }(-\log \lambda
_{n}^{k}-H(\rho _{0}))^{2}\lambda _{n}^{k}  \label{5-21}
\end{equation}%
\begin{equation}
=\log ^{2}\frac{N(\omega _{k})+1}{N(\omega _{k})}\sum_{n=0}^{\infty
}(n-N(\omega _{k}))^{2}\frac{N(\omega _{k})^{n}}{(N(\omega _{k})+1)^{n+1}}%
=F(N(\omega _{k})),
\end{equation}%
where
\begin{equation*}
F(x)=x(x+1)\log ^{2}{\frac{x+1}{x}}
\end{equation*}%
is a uniformly bounded function on $(0,\infty )$. Thus
\begin{equation}
\mbox{Tr}\rho _{0}(I-P_{0})\leq \frac{\sum_{k}F(N(\omega _{k}))}{\delta
^{2}T^{2}}\leq \frac{c_{1}s_{T}}{\delta ^{2}T^{2}}=\frac{c_{2}\bar{\omega}(T)%
}{\delta ^{2}T},  \label{5-23}
\end{equation}%
and a similar estimate holds for $\mbox{Tr}${$\rho $}$_{\pi }(I-P)$ with $%
N(\omega _{k})$ replaced by $N_{k}^{\prime }=N(\omega _{k})+|K(\omega
_{k})|^{2}m_{k}^{\ast }$.

Let \textsf{P} be a distribution on the set of $N$ \ \textquotedblleft
words\textquotedblright\ $x^{1},\dots ,x^{N}$, under which the words are
independent and have the probability distribution (\ref{4-7}). Let
\begin{equation*}
\nu _{T}=\mathsf{P}({\frac{1}{s_{T}}}\sum_{k=1}^{s_{T}}\hbar \omega
_{k}|x_{k}|^{2}\leq E),
\end{equation*}%
and remark that $\mathsf{E}{\frac{1}{s_{T}}}\sum_{k=1}^{s_{T}}\hbar \omega
_{k}|x_{k}|^{2}\leq E$ (where \textsf{E} is the expectation corresponding to
\textsf{P}), hence by the central limit theorem
\begin{equation*}
\lim_{T\rightarrow \infty }\nu _{T}\geq 1/2.
\end{equation*}%
Let us explain why the central limit theorem holds for sums $%
\sum_{k=1}^{s_{T}}\hbar \omega _{k}|x_{k}|^{2}$ as $T\rightarrow \infty .$
The summands are squares of the normal random variables $\xi _{k,T}=\sqrt{%
\hbar \omega _{k}}x_{k}$ which have zero means and the uniformly bounded
variances (see (\ref{ineq}))%
\begin{equation*}
\hbar \omega _{k}m_{k}^{\ast }\leq \frac{\hbar \omega _{k}}{|K(\omega
_{k})|^{2}}{\frac{1}{\mbox{e}^{c_{\infty}f(\omega _{k})}-1}=}c_{\infty}^{-1}%
\frac{c_{\infty}f(\omega _{k})}{\mbox{e}^{c_{0}f(\omega _{k})}-1}\leq
c_{\infty}^{-1}.
\end{equation*}%
Therefore the Liapunov condition is fulfilled ensuring convergence of the
properly normalized sums $\sum_{k=1}^{s_{T}}\xi _{k,T}^{2}$ to the normal
distribution.

Define the modified distribution $\mathsf{\tilde{P}}$ under which the words
are still independent but have the distribution
\begin{equation}
\tilde{\pi}_{\ast }(d^{2s_{T}}x)=\left\{
\begin{array}{ll}
\nu _{T}^{-1}\pi _{\ast }(d^{2s_{T}}x), & \mbox{if}\,\sum_{k=1}^{s_{T}}\hbar
\omega _{k}|x_{k}|^{2}\leq ET, \\
0, & \mbox{otherwise.}%
\end{array}%
\right.  \label{1a-11}
\end{equation}%
Therefore ${\tilde{\mathsf{E}}}\xi \leq \nu _{T}^{m}\mathsf{E}\xi \leq 3^{m}%
\mathsf{E}\xi $ for any nonnegative random variable $\xi $ depending on $m$
words and $T$ large enough.

Now let $x^{1},...,x^{N}$ be taken randomly with the joint probability
distribution $\tilde{\mathsf{P}}.$ Since the right hand side of (\ref{5-17})
depends at most on $m=2$ words,
\begin{equation*}
{\tilde{\mathsf{E}}}\inf_{M}{\bar{\lambda}}(\Sigma ,M)\leq \frac{1}{N}%
\sum_{j=1}^{N}\left\{ 9\mathsf{M}\mbox{Tr}\rho _{x^{(j)}}(I-P)+\mbox{Tr}\rho
_{0}(I-P_{0})+\sum_{k\neq j}9\mathsf{E}\mbox{Tr}P\rho
_{x^{(j)}}PP_{x^{(k)}}\right\}
\end{equation*}%
\begin{equation*}
=9\mbox{Tr}{\rho }_{\pi }(I-P)+\mbox{Tr}\rho _{0}(I-P_{0})+9(N-1)\mbox{e}%
^{-\left( C_{\chi ,T}-2\delta T\right) }
\end{equation*}%
\begin{equation*}
\leq \frac{c_{3}\bar{\omega}(T)}{\delta ^{2}T}+9\mbox{e}^{\left( RT+2\delta
T-C_{\chi ,T}\right) },
\end{equation*}%
where (\ref{5-23}) was used to estimate the first two terms. To complete the
proof we have only to show that
\begin{equation}
\liminf_{t\rightarrow \infty }\frac{C_{\chi ,T}}{T}\geq C.  \label{5-24}
\end{equation}

Let $m^{\ast }(\omega )$ be the function (\ref{5-15}), and let $\omega
_{k}^{\prime }$ be the point on the segment $[\omega _{k-1},\omega _{k}]$ at
which it achieves its minimum, then
\begin{equation*}
{\frac{1}{2\pi }}\hbar \omega _{k}^{\prime }m^{\ast }(\omega _{k}^{\prime
})\leq \int_{\underline{\omega }}^{\bar{\omega}}\hbar \omega m^{\ast
}(\omega ){\frac{d\omega }{2\pi }}=E,
\end{equation*}%
hence
\begin{equation*}
{\frac{C\chi ,_{T}}{T}}\geq \sum_{k=1}^{s_{T}}[g(N(\omega _{k})+|K(\omega
_{k})|^{2}m^{\ast }(\omega _{k}^{\prime }))-g(N(\omega _{k}))]{\frac{\Delta
\omega _{k}}{2\pi }}.
\end{equation*}%
Since $N(\omega ),$ $K(\omega )$ and $m^{\ast }(\omega )$ are continuous and
the summand is nonnegative, the limit of the last sum is greater than or
equals to
\begin{equation*}
\int_{0}^{\tilde{\omega}}[g(N(\omega )+|K(\omega )|^{2}m^{\ast }(\omega
))-g(N(\omega ))]d\omega ,
\end{equation*}%
for any fixed $\tilde{\omega}>0.$ Letting $\tilde{\omega}\uparrow \infty $
we obtain (\ref{5-24}) and the proof is completed.

\section*{Appendix 3. The infinite capacity}

We have seen that in the \textit{quantum broadband noise model} the capacity
is finite as follows from the estimate of the Theorem in Sec.2.

Let us show that the capacity can be infinite in the \textit{quantum
bandpass noise model} (Subsec. 3.2). For simplicity we consider the case $%
K(\omega )\equiv 1.$ Let $N(\omega )\geq 0$ be the spectral density of the
quantum noise and $m(\omega )$ a spectral distribution of the signal. Then
\begin{equation*}
C=\sup_{m\in \mathcal{M}}\int_{0}^{\infty }[g(N(\omega )+m(\omega
))-g(N(\omega ))]\frac{d\omega }{2\pi },
\end{equation*}%
where
\begin{equation*}
\mathcal{M}=\{m(\cdot ):m(\omega )\geq 0,\quad \hbar \Omega \int_{0}^{\infty
}m(\omega )\frac{d\omega }{2\pi }=E\}.
\end{equation*}%
Here $\Omega $ is the \textquotedblleft carrier frequency\textquotedblright
. Assume that $N(\omega )$ is monotonously decreasing for $\omega $ large
enough and tends to 0 as $\omega \rightarrow \infty .$ We will show that $%
C=\infty $ by choosing rectangular $m(\omega )$ such that%
\begin{equation*}
m(\omega )=\left\{
\begin{array}{cc}
M & \text{if }\omega \in \lbrack \omega _{1},\omega _{2}] \\
0 & \text{otherwise}%
\end{array}%
\right. ,
\end{equation*}%
where $M=\frac{2\pi E}{\hbar \Omega \left( \omega _{2}-\omega _{1}\right) }.$
We use the fact that if $\phi (x)$ is concave increasing function (so that $%
\phi ^{\prime }(x)$ is decreasing), then%
\begin{equation}
\phi (x+y)-\phi (x)\geq \phi ^{\prime }(x+y)\,y,\quad x,y\geq 0.  \label{aa}
\end{equation}%
Applying this for $\phi (x)=g(x)$ and using the fact that $g^{\prime
}(x)=\log (x+1)-\log x$ is decreasing function with $g^{\prime }(x)\geq
-\log x,$ we obtain%
\begin{eqnarray*}
C &\geq &\int_{\omega _{1}}^{\omega _{2}}[g(N(\omega )+M)-g(N(\omega ))]%
\frac{d\omega }{2\pi } \\
&\geq &\int_{\omega _{1}}^{\omega _{2}}g^{\prime }(N(\omega )+M)M\frac{%
d\omega }{2\pi } \\
&\geq &g^{\prime }(N(\omega _{1})+M)M\frac{\left( \omega _{2}-\omega
_{1}\right) }{2\pi } \\
&=&\frac{E}{\hbar \Omega }g^{\prime }(N(\omega _{1})+M) \\
&\geq &-\frac{E}{\hbar \Omega }\log (N(\omega _{1})+M).
\end{eqnarray*}%
Choosing $\omega _{1}\rightarrow \infty ,\,\omega _{2}-\omega
_{1}\rightarrow \infty ,$ which amounts to $\omega _{1}\rightarrow \infty
,M\rightarrow 0,$ we obtain $\,C=\infty .$

Similarly, in the \textit{classical case} we have%
\begin{equation*}
C=\frac{1}{2}\sup_{m\in \mathcal{M}}\int_{0}^{\infty }[\log (N(\omega
)+m(\omega ))-\log (N(\omega ))]\frac{d\omega }{2\pi },
\end{equation*}%
where
\begin{equation*}
\mathcal{M}=\{m(\cdot ):m(\omega )\geq 0,\quad \int_{0}^{\infty }m(\omega )%
\frac{d\omega }{2\pi }=E\}.
\end{equation*}%
Then applying (\ref{aa}) to $\phi (x)=\log \,x$ and using the fact that $%
\left[ \log \,x\right] ^{\prime }=x^{-1}$ is decreasing function we obtain
\bigskip
\begin{equation*}
C\geq \frac{1}{2}\frac{E}{N(\omega _{1})+M}\rightarrow \infty
\end{equation*}%
as $\omega _{1}\rightarrow \infty ,\,\omega _{2}-\omega _{1}\rightarrow
\infty .$ \bigskip

\textbf{Acknowledgments.} The author is grateful to G. De Palma, M. E.
Shirokov and D. Ding for comments improving the presentation. This work is supported by
the Russian Science Foundation under grant 14-21-00162.


\begin{thebibliography}{99}
\bibitem{bs} B. R. Bardhan, J. H. Shapiro, Ultimate capacity of linear
time-invariant Bosonic channel, Phys. Rev. A 93, 032342 (2016).


\bibitem{xcaves} C. M. Caves, P. B. Drummond, Quantum limits of bosonic
communication rates, Rev. Mod. Phys., vol. 66, no. 2, pp. 481-538
1994.

\bibitem{pmg} G. De Palma, A. Mari, V. Giovannetti, Classical capacity of
Gaussian thermal memory channels, Phys. Rev. A 90, 042312 (2014).


\bibitem{xgal} R. G. Gallager, \textsl{Information Theory and Reliable
Communications}. New York: J. Wiley 1968.

\bibitem{ghg} V. Giovannetti, A. S. Holevo, R. Garcia-Patr\'{o}n, A solution
of Gaussian optimizer conjecture for quantum channels, Communications in
Mathematical Physics, 334:3 (2015), 1553-1571.

\bibitem{glm} V. Giovannetti, S. Lloyd, L. Maccone, P.W. Shor, Broadband
channel capacities, Phys. Rev. A 68, 062323 (2003).



\bibitem{hall} M. J. W. Hall, Quantum information and correlation bounds,
Phys. Rev. A 55,(1997) 1050-2947.

\bibitem{xhe} C. W. Helstrom, \textsl{Quantum Detection Theory}. Progress in
Optics, vol. 10, 1972, 291-369.


\bibitem{xstat} A. S. Holevo, \textsl{Investigations in the General Theory
of Statistical Decisions}, Proc. of the Steklov Institute of Mathematics,
vol. 124, 1976 (AMS Translation 1978, Issue 3).

\bibitem{hol98} A. S. Holevo, Quantum coding theorems. Russian Math.
Surveys, vol. 53, N6, 1998, 1295-1331. Arxiv:quant-ph/9809023.

\bibitem{hh} A. S. Holevo, Single-mode quantum Gaussian channels: structure
and quantum capacity, Probl. Inform. Transmission, 43:1 (2007), 1--11.

\bibitem{hol11} A. S. Holevo, \textsl{Probabilistic and Statistical Aspects
of Quantum Theory, 2nd edition}, Edizioni della Normale, Pisa 2011.

\bibitem{hol} A. S. Holevo, Information capacity of quantum observable,
Probl. Inform. Transmission 48:1 (2012) 1-10.

\bibitem{hol13} A. S. Holevo, \textsl{Quantum systems, channels, information. A mathematical introduction}, De Gruyter, Berlin–Boston 2013. 

\bibitem{shir} A. S. Holevo, M. E. Shirokov, Continuous ensembles and the $%
\chi$-capacity of infinite dimensional channels. Probab. Theory and Appl.
vol. 50, N1, 2005, 98-114.

\bibitem{hw} A. S. Holevo, R. F. Werner, Evaluating capacities of Bosonic
Gaussian channels. Phys. Rev. A. , vol. 63, 2001, 032312.


\bibitem{xle} D. S. Lebedev, L. B. Levitin, The maximal amount of
information transmissible by an electromagnetic field, Information and Control, vol.
9, 1966, 1-22.

\bibitem{man} F. Manuceau, A. Verbeure, Quasi-free states of the CCR algebra
and Bogoliubov transformations, Commun. Math. Phys. 9 (1968), 293-302.



\bibitem{hp} K. R. Parthasarathy, An introduction to quantum stochastic
calculus, Birkh\"auser Verlag, Basel-Boston-Berlin 1992.

\bibitem{sl} L. N. Slobodeckij, Generalized Sobolev spaces and their
applications to boundary value problems of partial differential equations,
Leningrad. Gos. Ped. Inst. U\v{c}ep. Zap. 197 (1958), 54--112.

\end{thebibliography}
\end{document}